\documentclass[12pt]{iopart}
\usepackage{graphicx}
\usepackage{epstopdf}
\usepackage{breqn}
\usepackage{color}
\usepackage{cite}
\renewcommand{\figurename}{Figure}

\begin{document}

\title[Particle - In - Cell observations of Brillouin scattering in  magnetized  plasma]
{Particle - In - Cell observations of Brillouin scattering for laser interacting with  magnetized overdense plasma}

\author{Laxman Prasad Goswami$^{*1}$, Rohit Juneja$^*$, Dhalia Trishul$^*$, Srimanta Maity$^*$, Sathi Das$^*$ and Amita Das$^{*2}$}

\address{$^{*}$Department of Physics, Indian Institute of Technology Delhi, Hauz Khas, New Delhi-110016, India \\}
\ead{$^{1}$goswami.laxman@gmail.com, $^2$amita@iitd.ac.in}
\vspace{10pt}

\begin{abstract}
 One dimensional Particle-in-cell simulations using   OSIRIS-4.0 has been  conducted  to  study the interaction of a laser electromagnetic pulse with  an overdense magnetized plasma target. The  external magnetic field has been chosen to be directed along the laser propagation direction. This geometry supports the propagation of right (R)  and left (L) circularly polarised electromagnetic waves in the plasma. The laser pulse is allowed to propagate inside the plasma when its frequency falls in the pass band of the dispersion curves of L and/or R waves. The strength of the applied external magnetic field is chosen as a  parameter to ensure  that the laser frequency lies in the appropriate pass band. It is demonstrated  that for all possible polarization of the 
 incident laser,  parametric process involving a scattered Electromagnetic wave and an electrostatic mode occur. The parametric process  has been identified as that due to the Brillouin back scattering process. 
\end{abstract}

\section{Introduction}
\label{sec:Introduction}
In recent years there has been a lot of progress (both fundamental and technological) in the area of laser plasma interaction studies \cite{das2020laser}. 
The development of low frequency pulsed $CO_2$ lasers \cite{tochitsky2016prospects, haberberger2010fifteen, beg1997study} and the possibility of generating  strong magnetic fields 
(e.g. of the order of Kilo Tesla has already been achieved \cite{nakamura2018record} and there are proposals to generate Mega Tesla \cite{korneev2015gigagauss} fields) 
are two  developments which have opened up possibilities of carrying out experiments in a new  direction 
involving  laser interacting with magnetized plasma. This regime is of importance   for  fundamental  explorations  and also  has rich  implications for   
 frontier applications such as direct heating of ions and neutron production in table top devices\cite{goswami2021ponderomotive, vashistha2020new, vashistha2021excitation, maity2021harmonic, mandal2021transparency, kumar2019excitation, mandal2020spontaneous, sano2020thermonuclear, sano2019ultrafast}. 
 In this paper we explore the prospect of  parametric scattering process associated with a  
 laser pulse interacting and propagating inside a magnetized plasma. The applied external magnetic field has been chosen to be along the laser propagation direction. This particular geometry   supports  the 
 propagation of Right (R) and Left (L) circularly polarised electromagnetic waves inside the plasma for the specific 
 range of pass band frequencies in their dispersion curves. 
 
The  parametric processes  have  been studied  extensively for the past several  decades. A detailed discussion of the same in the context of plasmas and its relevance to nuclear fusion experiments can be found in a recent review article \cite{kaw2017nonlinear}.  The parametric instability occurs when a   high-amplitude pump electromagnetic wave  couples with an electrostatic disturbance  in the plasma and generates a scattered  Electromagnetic radiation. The interaction of the pump and the scattered radiation accentuates 
the density disturbance and enhances the electrostatic perturbations, thereby creating a feedback loop for the instability to occur. Some early work in this area are as follows. Drake et al. \cite{drake1974parametric}, derived a compact dispersion relation for the wave interactions in the context of  an unmagnetized plasma. They also discussed the limiting form of the dispersion relation leading to the  various instabilities like Brillouin, Compton, and Raman scattering processes in plasma. In Raman scattering the excited electrostatic mode is an electron plasma wave, whereas for the  Brillouin scattering process it is essentially an ion acoustic wave. The analytical expression for the growth rate 
 and the threshold condition for the Raman and Brillouin scattering instability has been obtained by Liu et al. \cite{liu1974raman}. The occurrence of these Raman and Brillouin instabilities were directly shown for pulsed electromagnetic solitonic structures propagating in plasmas by 
Saxena et al. \cite{saxena2007stability} and Sundar et al. \cite{sundar2011relativistic} respectively 
with the help of fluid simulations.   The application of Brillouin scattering process  in the context of laser fusion 
has been highlighted in many works \cite{kruer1973instability, tsytovich1973one, weiland1977coherent, panwar2009stimulated}. 
For magnetized plasma theoretical studies have been carried out for the parametric instability by many authors. For instance,  Stenflo \cite{stenflo1990stimulated, stenflo1995theory} and Shukla \cite{shukla2010stimulated} 
provided a theoretical  prediction of ionospheric heating experiments. Jaiman and Tripathi \cite{jaiman1998stimulated} analyzed the generation of Brillouin and Compton scattering in magnetized plasma for oblique propagation  of  electromagnetic pump wave with respect to the external magnetic field direction. 

We provide here an evidence of parametric process occurring for an electromagnetic pulse 
propagating inside a   magnetized plasma using 
particle-in-cell (PIC) simulations \cite{dawson1983particle, birdsall1991particle}. 
The laser is chosen to be incident on an overdense plasma and it propagates parallel to the applied  external   magnetic field. This particular geometry supports the propagation of Right (R) and/or Left (L)  circularly polarised  
electromagnetic (EM) waves even for an  overdense plasma,   provided the EM wave frequency lies in the appropriate pass band of their dispersion relation. 

The manuscript has been  organized as follows. Section \ref{sec:SimulationDetails} provides  simulation details. In section \ref{sec:Observations}, we present  the details of  observations made by simulating 
different cases of  polarization of the incident EM pulse. These observations have been analyzed in detail 
in Section{\ref{sec:CharacterizationOfModes}} which provide evidence of the occurrence of parametric 
 Brillouin back scattering process. In a recent work from our group \cite{goswami2021ponderomotive} we had 
 observed the generation of electrostatic fluctuations for  a sharp profile of the laser pulse in the same geometry by the ponderomotive pressure 
 of the laser pulse. Here, in contrast for a smoother laser profile, we observe electrostatic fluctuations getting generated by the Brillouin backscattering process. We point out in section\ref{sec:LaserProfiles} that both these processes in fact take place. The ponderomotive pressure driven process is the first to occur and gets saturated if the profile is relatively smooth. However the fluctuations generated from this process 
 mimic the effective temperature of electron fluid to drive the brillouin scattering process which occurs at 
 a later stage.  We summarize and conclude in section \ref{sec:Conclusion}. 

\section{Simulation details}
\label{sec:SimulationDetails}

\begin{figure*}
	\centering
	\includegraphics[width=6.0in]{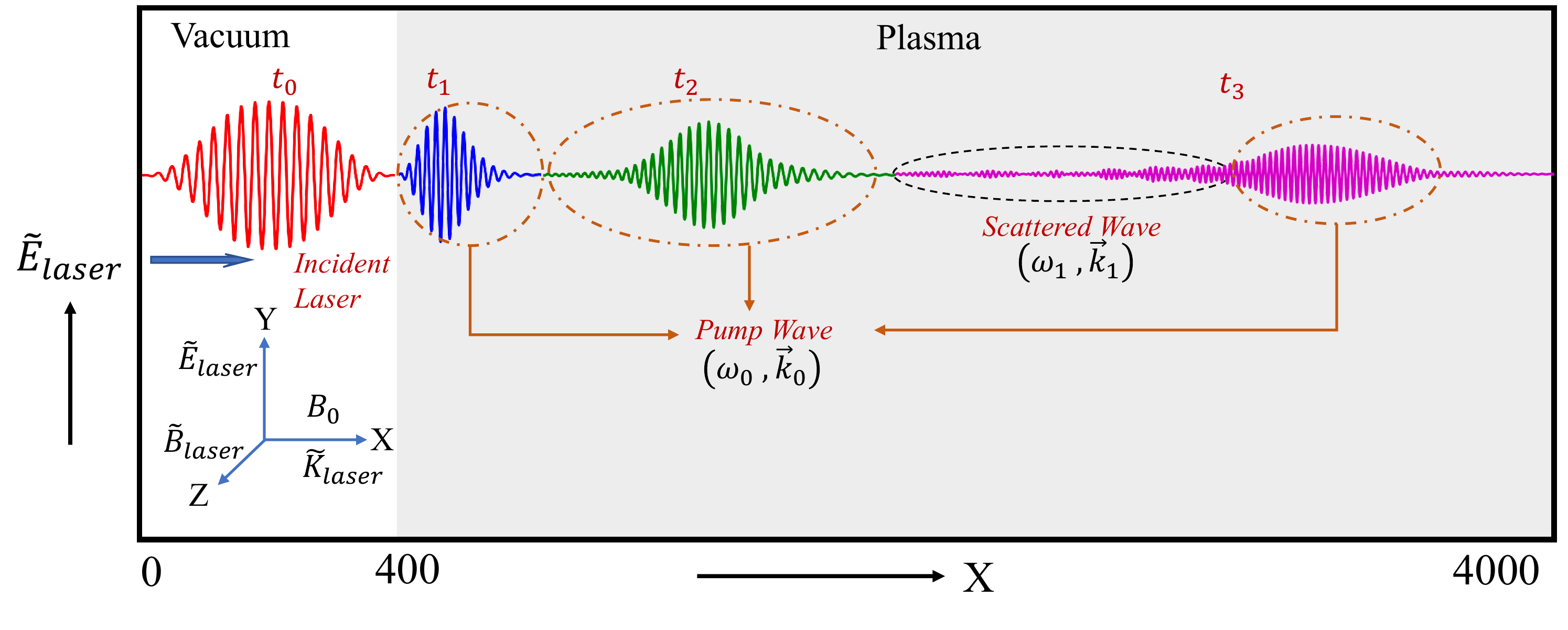}
	\caption{Schematics (not to scale) shows the simulation setup and a summary of the physical process observed in our study. We have performed 1D particle-in-cell (PIC) simulations using OSIRIS. The laser is propagating along the X-direction. The laser is incident at time $t_0$ from the left side of the simulation box on the vacuum-plasma interface at $x = 400$. We have chosen the RL-mode geometry, where the external magnetic field $(B_0)$ is applied along the laser propagation direction. Here $t_0$, $t_1$, $t_2$, and  $t_3$ represent different simulation  times (in ascending order).  It is evident from the schematic that there is an excitation of electromagnetic (EM) mode (pump wave) in the plasma at time $t_1$. The pump wave broadens with its propagation inside the plasma at time $t_2$. With further propagation of pump wave, scattered waves are excited in the bulk of plasma at time $t_3$. The amplitude of these scattered radiations increases with time.}
	\label{fig:SchematicBrillouin}
\end{figure*}

We have employed the OSIRIS-4.0 framework \cite{hemker2000particle, fonseca2002osiris, fonseca2008one} for carrying out 1-D particle-in-cell (PIC) simulations to study the interaction of a laser with magnetized plasma. 
The schematic of the simulation geometry (not to scale) has been shown in Fig.\ref{fig:SchematicBrillouin}. The external magnetic field is directed along the laser propagation direction $\hat{x}$. A 1-D simulation box with dimension $L_{x} = 4000 c/\omega_{pe}$ has been chosen. Here $c$ is the velocity of light and $\omega_{pe}$ represents the 
plasma frequency. 
The number of particles per cell are taken to be $8$. The plasma boundary starts from $x = 400 c/\omega_{pe}$. There is vacuum region between $x = 0$ to $ 400 c/\omega_{pe}$ with a sharp plasma vacuum interface at $x=400 c/\omega_{pe}$. 
The spatial resolution is taken as $100$ cells per electron skin depth and it  corresponds to the grid size $\Delta x = 0.01 c/\omega_{pe}$. The laser is incident on the plasma target from left side. We consider a short-pulse laser of 
frequency $\omega_l = 0.3 \omega_{pe}$.  
The laser profile is Gaussian having rise and fall time of $200 \omega_{pe}^{-1}$ ($400 \omega_{pe}^{-1}$ for some simulations) with the peak intensity of $I = 3.5 \times 10^{19} Wm^{-2}$ (corresponding to a relativistic factor $a_0 = 0.5$). Boundary conditions are taken as absorbing in the longitudinal direction. The PIC code OSIRIS uses normalised values of the various parameters. 
We have followed the dynamics of both electrons and ions. To reduce the computational time, we carried out the simulations for  a reduced mass of ions, which is chosen  to be $25$ ( the value of $40$, $50$ are also chosen 
for some simulations) times heavier than electrons. Thus   $m_{i} = 25m_{e}$ ($40m_e$, $50m_e$ for some simulations),   where $m_{i}$ and $m_{e}$ represent the rest mass of the ion and electron species respectively. We have provided the laser and plasma simulation parameters both in normalized units alongside their  typical possible values in 
standard units in Table-{\ref{table:simulationtable}}. As per the convention the  frequencies have been normalised by  electron plasma frequency ($\omega_{pe}$), length by electron skin depth ($c/\omega_{pe}$), and the  electric and magnetic fields by  $m_ec\omega_{pe}e^{-1}$. 

\begin{table}
	\caption{Values of simulation parameters in normalized and standard units}
	\label{table:simulationtable}
	\begin{center}
		\begin{tabular}{|c|c|c|}
			\hline		
			\color{red}Parameters & \color{red}Normalized value & \color{red}Values in SI unit\\
			\hline
			\hline	
			\multicolumn{3}{|c|}{\color{blue}Plasma Parameters} \\
			\hline
			$n_0$ & $1.0$ & $1.34\times10^{20} cm^{-3}$ \\
			\hline
			$\omega_{pe}$ & $1.0$ & $0.67\times10^{15} rad/s$\\
			\hline
			$\omega_{pi}$ ($M/m = 25$) & $0.2$ & $0.13\times10^{15} rad/s$\\
			\hline
			\multicolumn{3}{|c|}{\color{blue}Laser Parameters} \\
			\hline
			$\omega_{l}$ & $0.30$ & $0.2 \times10^{15} rad/s$\\
			\hline
			$\lambda_{l}$ & $21$ & $9.42 \mu m$\\
			\hline
			Intensity &  $a_0 = 0.5$ &$3.5\times 10^{19} W/m^2$\\
			\hline
		\end{tabular}	
	\end{center}
\end{table}

We have considered different configurations of incident laser polarization by appropriately choosing the parameter $\alpha_i$ (it is $0$ for linear, and $\pm 1$ for right (RCP), left (LCP) circular polarizations respectively).  Here the incident and transmitted laser electric field is denoted as $\vec{E}_i = \tilde{E}_{i}\left(\hat{y} + i\alpha_i\hat{z}\right)e^{-i\omega t}$ and $\vec{E}_t = \tilde{E}_{t}\left(\hat{y} + i\alpha_t\hat{z}\right)e^{-i\omega t}$ respectively. Where $\alpha_i$ and $\alpha_t$ respectively denote the polarization of incident and transmitted EM pulse. The applied external  magnetic field $B_0$ in normalized units has been varied from $2.5$ to  $10$ for our simulation studies as tabulated in Table-\ref{table:simulationCases}. 

The four different cases of study carried out by us correspond to  four different choices of the incident laser EM field parameters. 
For  case (A)  and case (B), a linearly polarized laser ($\alpha_i = 0$) propagating along the $X$-direction is considered.
The two choices of magnetic field ensures that for  the given laser frequency and the choice of $B_0 = 2.5$ in case(A)  only  the  $R$ mode lies in the pass band of the dispersion curve, while for higher value of  $B_0 = 10$ in case(B) both $R$ and $L$  EM modes are in the pass band.   
The polarization of laser is taken RCP ($\alpha_i = +1$) and LCP ($\alpha_i = -1$) in case (C)  and (D) respectively.   
The transmitted wave also has the same polarization as the incident wave in these two cases  as listed in  Table -\ref{table:simulationCases}. 
A fifth and sixth possibility of choosing  LCP for incident EM wave with an external magnetic field of $B_0 = 2.5$ 
and RCP for $B_0 = 10$ respectively, does not permit any   EM wave propagation inside the plasma, and hence have not been reported. In Table-\ref{table:simulationCases}, $\omega_{cs} = \frac{q_sB_0}{m_s}$ is the gyro- frequency of the species $s$ 
of the plasma (where $s = e,i$ correspond to electron and ion  particles) in the  external magnetic field $B_0$.  

\begin{table}
	\caption{List of simulation cases, incident and transmitted laser polarization, and external magnetic field parameters}
	\label{table:simulationCases}
	\begin{center}
		\begin{tabular}{|c|c|c|c|c|c|}
			\hline		
			\color{red}Cases & \color{red}Incident Laser  & \multicolumn{3}{|c|}{\color{red}Magnetic Field Parameters} & \color{red}EM in Plasma\\ 
			\hline
			& \color{blue}$\alpha_i$ & \color{blue}$B_0$ &\color{blue}$\omega_{ce}$ & \color{blue}$\omega_{ci}$ & \color{blue}$\alpha_t$\\
			\hline
			\hline	
			\color{blue} Case A & Linear $\alpha_i = 0$ & $B_0 = 2.5$ & $2.5$ & $0.1$ & R-wave $\alpha_t = +1$ \\
			\hline
			\color{blue} Case B & Linear $\alpha = 0$ & $B_0 = 10$ & $10$ & $0.4$& R and L-waves $\alpha_t = \pm 1$ \\
						\hline
			\color{blue} Case C & RCP $\alpha = +1$ & $B_0 = 2.5$ & $2.5$ & $0.1$& R-wave $\alpha_t = +1$ \\
			\hline
			\color{blue} Case D & LCP $\alpha = -1$ & $B_0 = 10$ & $10$ & $0.4$& L-wave $\alpha_t = -1$ \\
			\hline
			
		\end{tabular}	
	\end{center}
\end{table}

\section{Observations}
\label{sec:Observations}
In a recent publication Goswami et al.  
 \cite{goswami2021ponderomotive} have employed similar geometry to study the interaction of a laser pulse with a magnetized plasma. The study had interestingly shown excitation of electrostatic perturbations at the electron plasma frequency. Since the background plasma is overdense, such a perturbation cannot be understood through a parametric excitation process. Instead it was shown that the difference in the ponderomotive pressure experienced by electrons and ions led to the charge separation leading to plasma oscillations.  Here, on the other hand we demonstrate  that with a choice of a  smoother 
 laser pulse profile (consequently weakening the ponderomotive pressure term) one can excite 
 parametric instability. Since the plasma is overdense  only the Brillouin scattering process is observed. 
 
 \begin{figure*}
	\centering
	\includegraphics[width=6.1in]{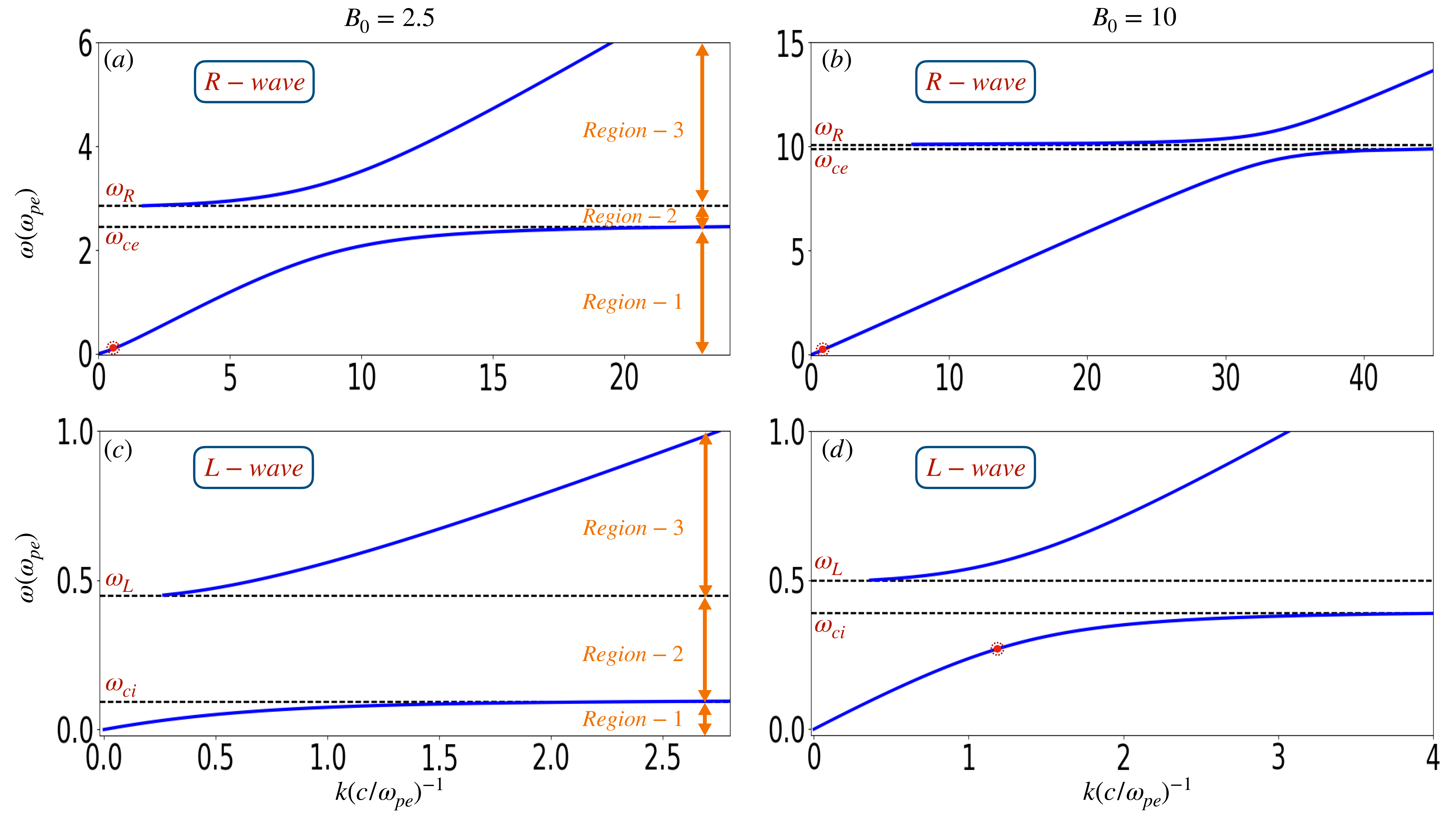}
	\caption{Dispersion curves for R-mode (a, b) and L-mode (c, d). Here $B_0$ is the normalized magnitude of external magnetic field applied. Fig. (a), (c) are for $B_0 = 2.5$ and Fig. (b), (d) are for $B_0 = 10$. Region-1 and Region-3 in each subplot represent the passband of the dispersion relation. Stopband is represented by the Region-2. Encircled red-dot in each figure represent the incident laser frequency. Laser frequency $\omega_l = 0.3\omega_{pe}$ lies in the passband of R-wave for both $B_0 = 2.5$ and $B_0 = 10$. For L-wave, $\omega_l = 0.3\omega_{pe}$ lies in the stopband for $B_0 = 2.5$ and passband for $B_0 = 10$.}
	\label{fig:DispersionCurve}
\end{figure*}

\begin{figure*}
	\centering
	\includegraphics[width=6.0in]{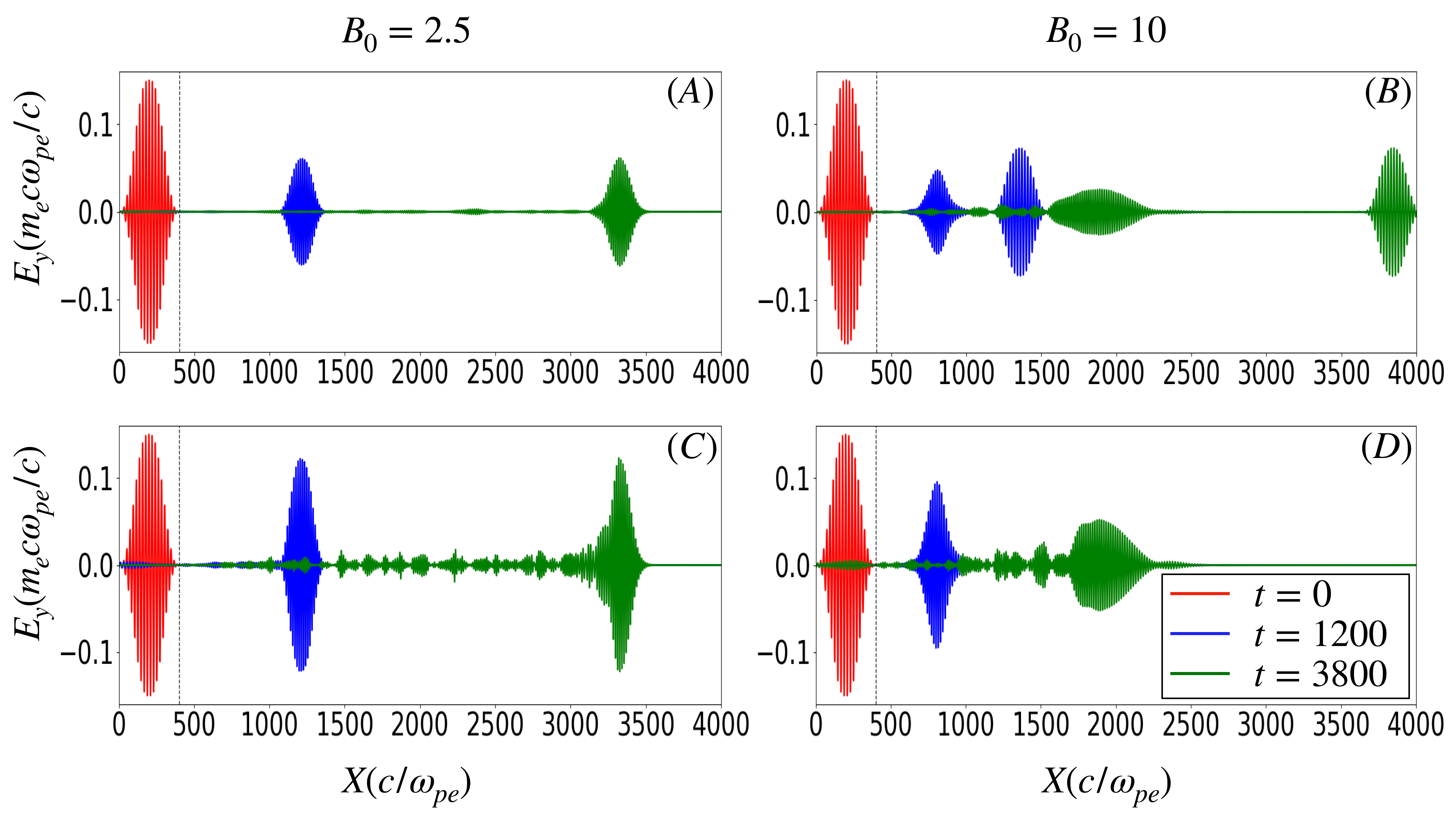}
	\caption{Figure shows the time evolution of the $y$ component of the EM wave inside the plasma for different values of the external magnetic field $B_0$ and the polarization of incident laser pulse for cases (A, B, C, D). At $t=0$, the EM wave is incident at vacuum-plasma boundary ($x=400$). The pump wave broadens with propagation in the bulk of plasma ($t = 1200$). Further propagation leads to side band EM scattering ($t=3000$).}
	\label{fig:EyPlots}
\end{figure*}
 
 We show in Fig.\ref{fig:DispersionCurve} the dispersion curves for both $R$ and $L$ waves for the two values of the magnetic 
 field $2.5$ and $10$ respectively. The incident laser frequency of $\omega_l = 0.3 \omega_{pe}$ 
 lies in the pass band of only the $R$ wave for the case of  $B_0 = 2.5$. For the other case of 
 $B_0 = 10$ the laser  frequency lies in the pass band of both $R$ and $L$ waves. The two waves, however, have distinct phase and group speeds.  
 The various subplots (A, B, C, D) of Fig.\ref{fig:EyPlots} correspond to the four different cases (A, B, C, D) respectively of study  listed in Table.2. 
 The $y$ component of the electric field  $E_y$  has been shown  at three different 
 times in this figure as a function of $x$. 
 While in subplot(A) and (B) the incident pulse has linear polarization 
 for (C) and (D) the incident pulse has been chosen to have Right and left circular polarization. 
 The pulse in  red color shows  $E_y$ for the 
 incident EM pulse at $t = 0$. It is placed in the vacuum region initially. The plasma starts from 
 $x= 400c/\omega_{pe}$. The blue and green pulses in the figure depict the $E_y$ component of electric field at $t = 1200 \omega_{pe}^{-1}$ and $t = 3800 \omega_{pe}^{-1}$ respectively. 
 For $B_0 = 2.5 $ both linear and right hand circular  polarised incident laser, shown in (A) and (C) 
 subplots propagate as $R$ waves inside the plasma. This is so because the laser frequency lies in the stop band of $L$ wave. The left circularly polarised incident radiation in this case gets 
 totally reflected for this particular case. For case (B) the incident linearly polarised EM pulse generates both $L$ and $R$ waves, as the frequency lies inside the pass bands of both the modes. The $L$ and $R$ waves get  spatially separated as they propagate inside the plasma due to their different group speeds. The $R$ wave being faster it moves ahead. For case (D), the incident laser pulse is chosen to have left circular polarization. This pulse propagates as $L$ wave inside the plasma. Thus Case (A) and (C) show the propagation of pure $R$ waves inside the plasma and case (D) shows pure $L$ wave propagation. Case(B) on the other hand depicts the propagation of both $L$ and $R$ wave.  Thus for these cases one can study separately  the propagation  of each of the $R$ and $L$ waves without getting hindered by the other wave in any fashion.  

 \begin{figure*}
	\centering
	\includegraphics[width=6.0in]{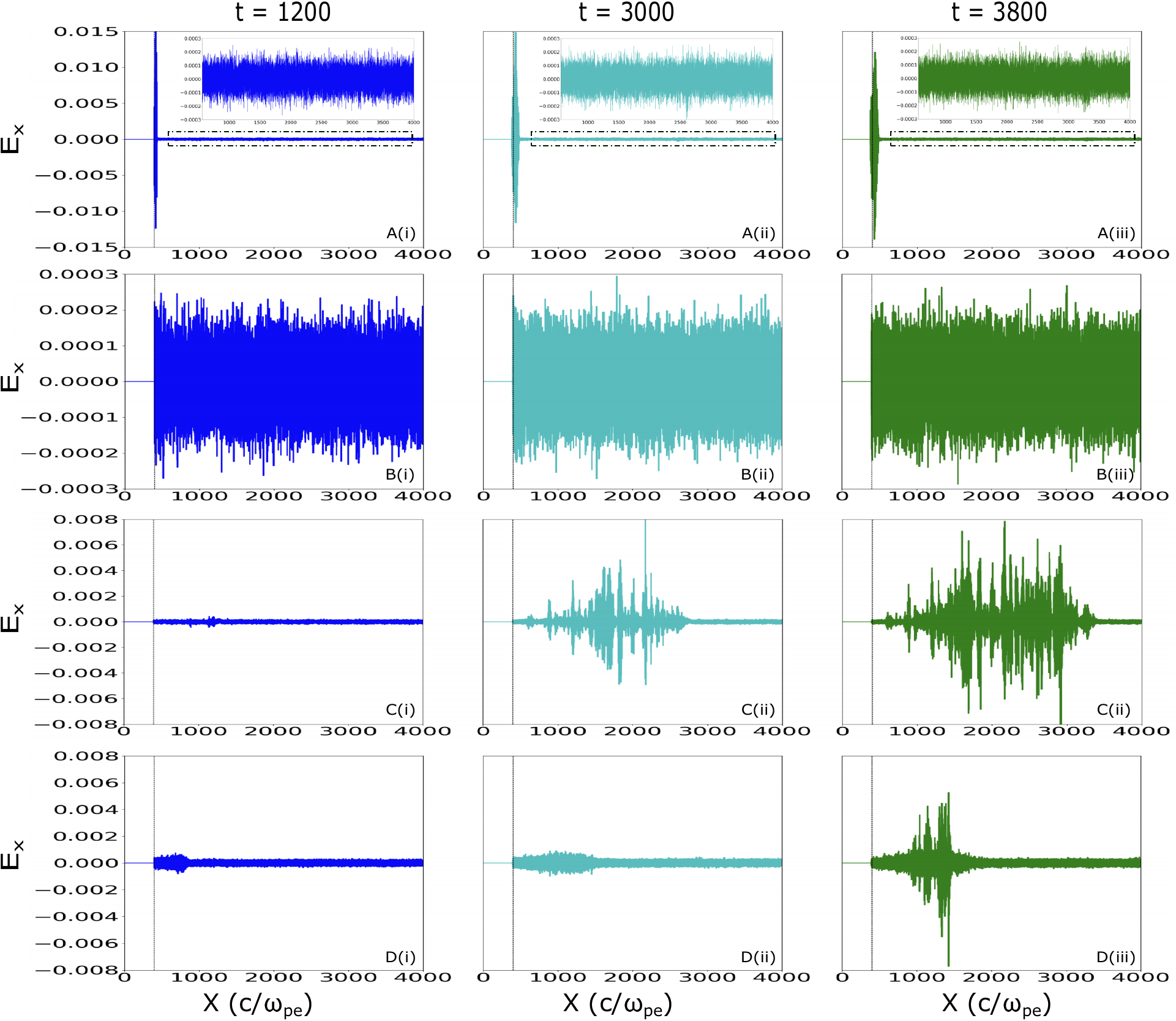}
	\caption{The figure shows the time evolution of the $x$ component of the electric field inside the plasma for cases (A, B, C, D). Initially the amplitude of $E_x$ is small ($t=1200$). It is evident from the figure that as the Brillouin scattering starts, laser amplitude electrostatic fluctuations are observed in the bluk of plasma ($t=3000$, and  $t=3800$).}
	\label{fig:ExPlots}
\end{figure*}

\begin{figure*}
	\centering
	\includegraphics[width=6.0in]{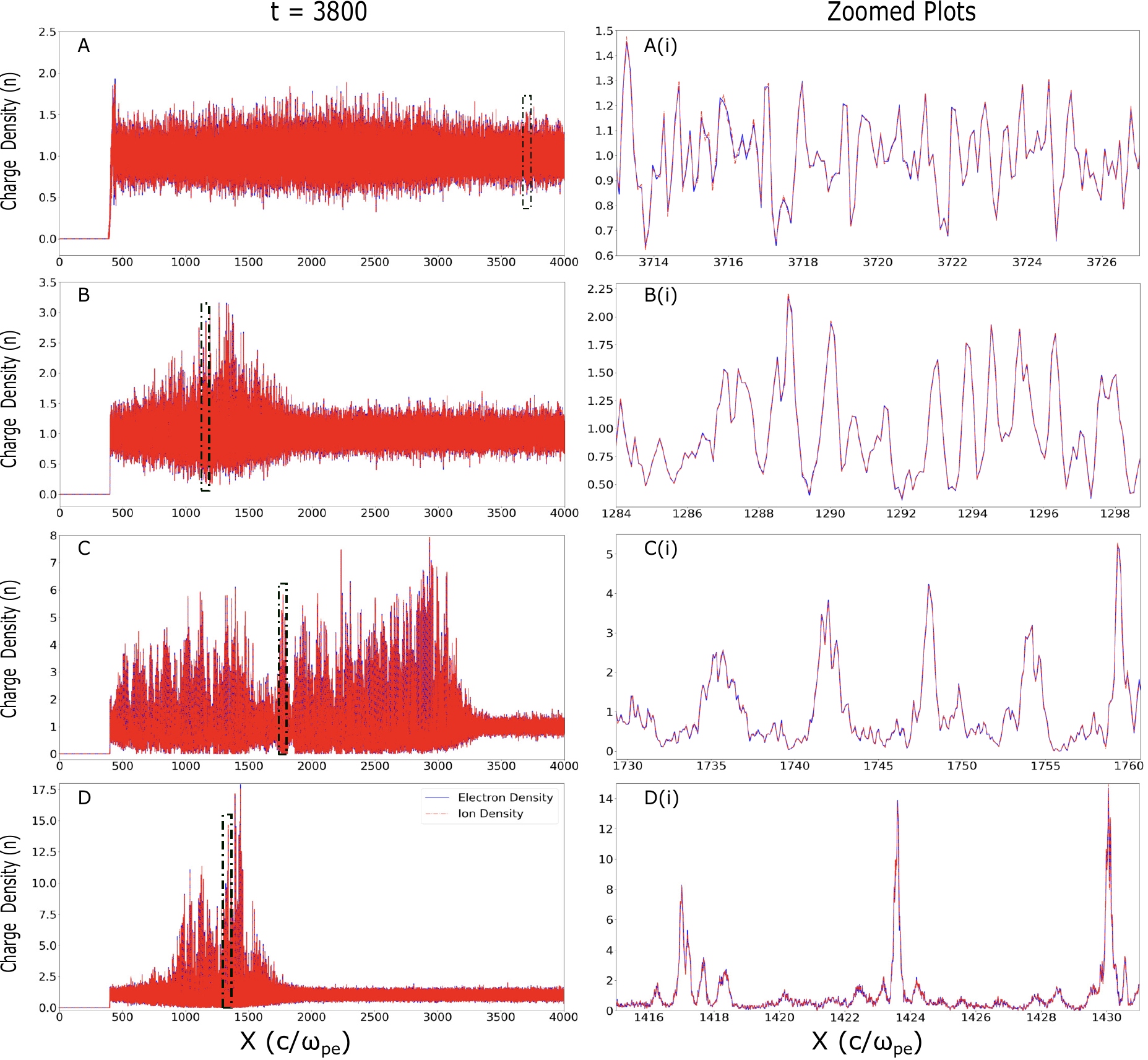}
	\caption{The figure shows the electron ($n_e$) and ion ($n_i$) density fluctuations at $t = 3800$ for the cases (A, B, C, D). The zoomed plot alongside  indicates that electrons and ions density perturbations have similar form. This is an evidence of the Brillouin scattering phenomenon.}
	\label{fig:DensPlots}
\end{figure*}
 
 From Fig.\ref{fig:EyPlots} it is clear that as the  EM pulse propagates inside the plasma, electrostatic field 
 disturbances get generated 
 behind the pulse. These disturbances have a higher amplitude  when the  incident laser pulse 
 has circular polarisation. This may be so as the power level of the propagating $L$ and $R$ waves for the linear case get low compared to the case of circular polarization. Associated with these electromagnetic disturbances one can observe electrostatic perturbation which are shown as the plot of $E_x$ in Fig.\ref{fig:ExPlots} at three different times. The corresponding electron and ion charge density perturbations are shown in Fig.\ref{fig:DensPlots} for $t = 3800$. The zoomed plot alongside shows that the electron (in solid blue line) and ion (in red dash line) density perturbations have similar form. In the next section 
 we try to characterize these perturbations and provide the evidence that they represent the Brillouin scattering phenomena.

\section{Characterization of the observed excitation as a Brillouin process }
\label{sec:CharacterizationOfModes}

\begin{figure*}
	\centering
	\includegraphics[width=6.0in]{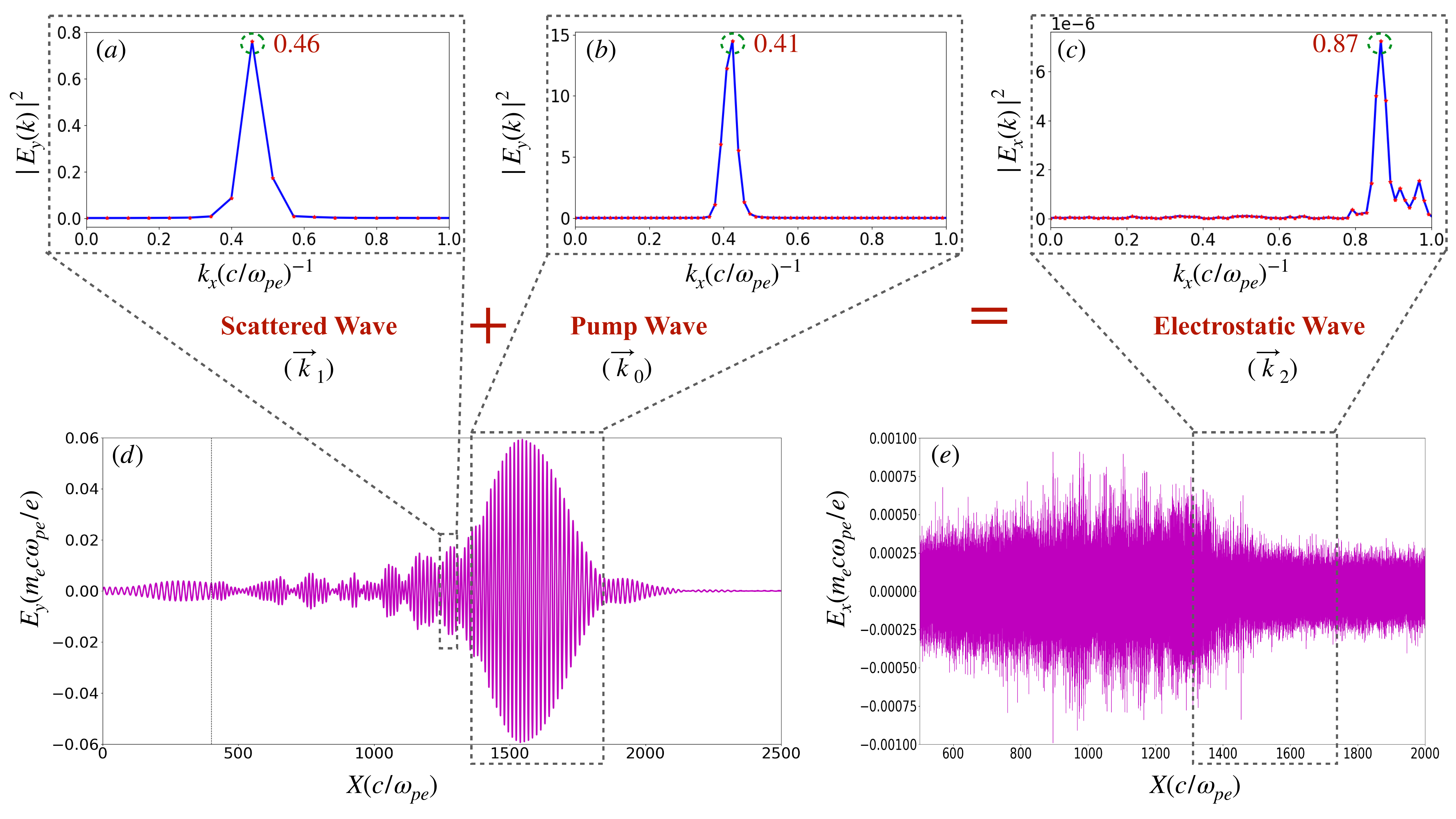}
	\caption{Figure gives the space Fast Fourier Transform (FFT) of (a) scattered wave, (b) pump wave, and (c) the electrostatic (ES) modes (c) for case C. In this case, the incident laser pulse is left circularly polarized (LCP). The FFT is performed at time $t = 3000$ in the bulk of plasma, when there is generation of scattered waves and ES modes.}
	\label{fig:SpaceTimeFFTLCPParametricInstability}
\end{figure*}

We evaluate the spatial Fourier transform (FFT) spectra for the electromagnetic $E_y$ and electrostatic $E_x$ fluctuations. The spectral wavenumber peak for one particular case (D) of the $L$ wave propagation has been shown in Fig.\ref{fig:SpaceTimeFFTLCPParametricInstability}.  The FFT of $E_y$ field using only the region of  pump pulse region shows a peak at $k_0 = 0.41$ (which satisfies the Dispersion curve for the $L$ wave. However, when FFT is taken of the $E_y$ in a region just behind the main pulse the peak of the spectra occurs slightly shifted at $k_1=0.46$. This appears to be the scattered EM radiation. The fourier spectra of the electrostatic field $E_x$  on the other hand peaks at $k_2 = 0.87$. It is interesting to note that the wave-vector matching condition for the parametric process is satisfied as $\vec{k_2} = \vec{k_0} + \vec{k_1}$. The electrostatic perturbation being at shorter scales than the pump and scattered EM radiation shows that it is a back scattering process that is taking place. Furthermore, as shown in Fig.\ref{fig:DensPlots}, the electron and ion density perturbations seem to occur in phase indicating that it is a Brillouin process. Furthermore, we have also repeated these simulations for the case of static ions (corresponding to $m_i = \infty$). For such a case no scattering is observed.

The simulations were carried out for the case of a plasma which was cold. The question then  arises is what plays   the role of  effective temperature for exciting an ion acoustic perturbation in the Brillouin process. We feel that the velocity acquired by the electrons as a result of ponderomotive pressure  from the EM pulse may contribute for the same. In fact similar observations were noted in the work by Sundar et. al. \cite{sundar2011relativistic} where they observed a Brillouin scattering process for a flat top solitonic structure in a fluid simulation. We will discuss more about the role the  
ponderomotive pressure term plays by choosing different laser profiles in section \ref{sec:LaserProfiles}.

\begin{figure*}
	\centering
	\includegraphics[width=5.0in]{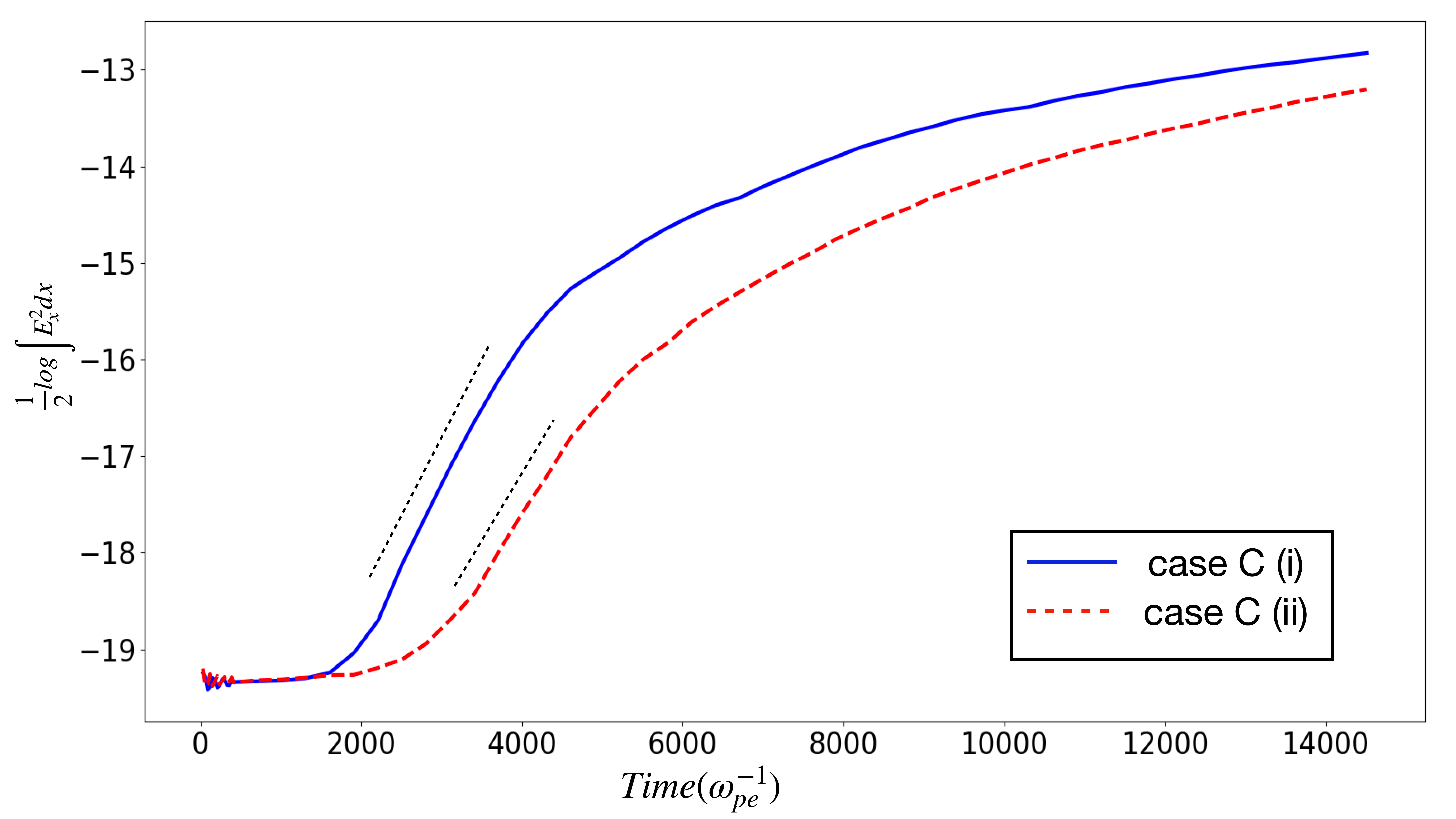}	
	\caption{Figure shows the time evolution of electrostatic energy for (i) $m_i = 25$ and (ii) $m_i = 50$ for case C, when the incident laser pulse is right circularly polarized (RCP). The growth is calculated using the slope of $\log\left(\frac{E_x^2}{2}\right)$ vs. time plot. For each curve, the slope is taken along the black-dash line.}
	\label{fig:GrowthRateRCP}
\end{figure*}

\begin{figure*}
	\centering
	\includegraphics[width=5.0in]{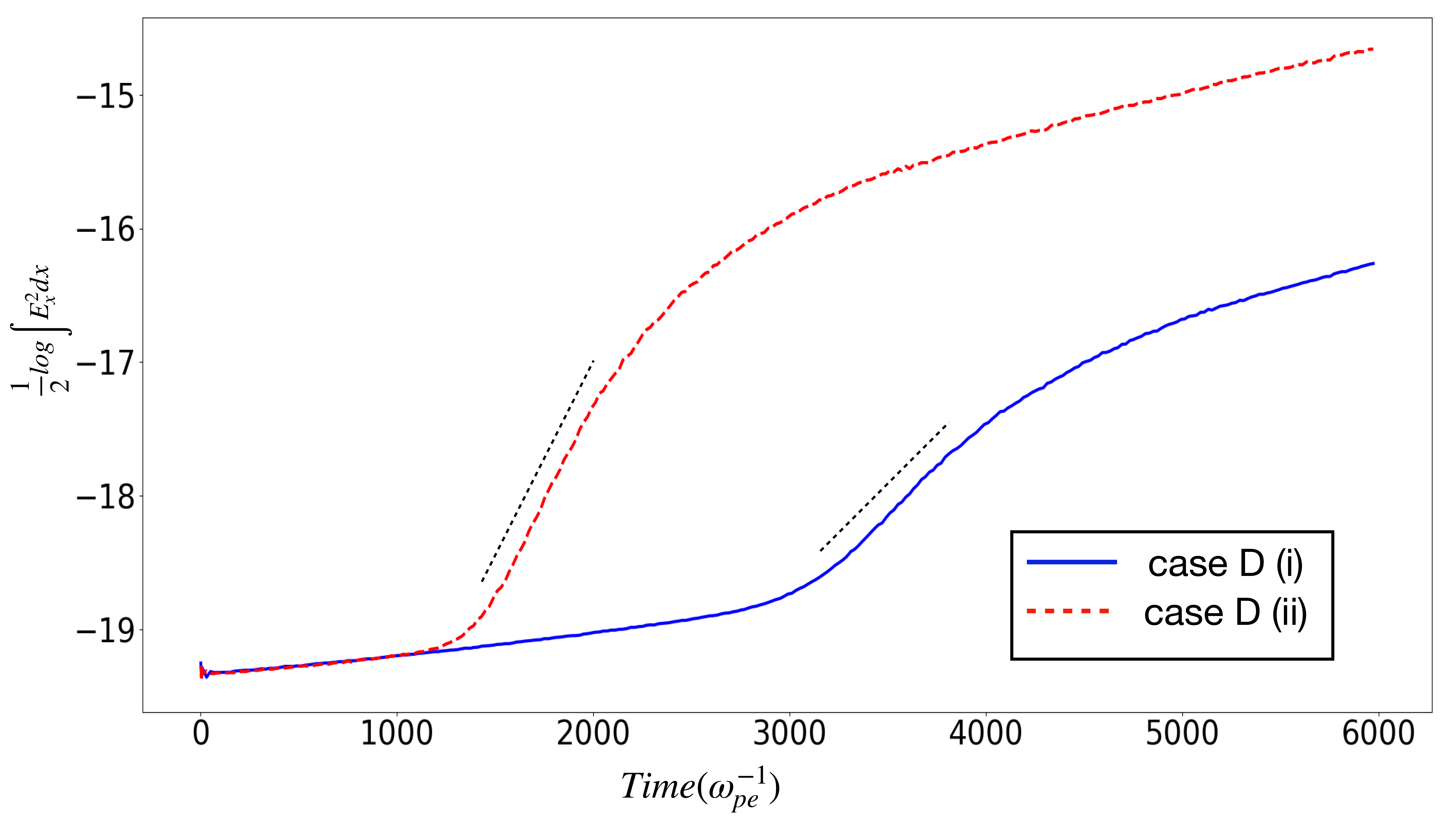}	
	\caption{Figure shows the time evolution of electrostatic energy for (i) $m_i = 40$ and (ii) $m_i = 50$ for case D, when the incident laser pulse is left circularly polarized (LCP). The growth is calculated using the slope of $\log\left(\frac{E_x^2}{2}\right)$ vs. time plot. For each curve, the slope is taken along the black-dash line.}
	\label{fig:GrowthRateLCP}
\end{figure*}

In Fig.\ref{fig:GrowthRateRCP} we have shown the evolution of electrostatic field energy for Case(C) of $R$ wave propagation for two different ion to electron mass ratios (e.g. $25 m_e$ and $50 m_e$). The other parameters for the laser pulse and the plasma medium  are the same. It is interesting to observe that the initial slope for the two cases are identical. Subsequently a second phase with a faster growth rate occurs in both the cases. The onset of second phase occurs at an earlier  time for ion mass $m_i = 25m_e$ compared to that of $m_i = 50m_e$. Also, the growth rate as discerned from the slope in this phase for the 
lower ion mass is higher. This can be understood by realising that the initial small rise of the  electrostatic energy arises as a result of ponderomotive forcing from the laser pulse, during which the electrons acquire a certain kinetic energy which acts as an effective temperature for facilitating the Brillouin scattering process. The second phase corresponds to the Brillouin scattering instability during which the two growth rate differs. After the second phase the nonlinear effects seem to set in which slows down the growth rate. 
In Fig.\ref{fig:GrowthRateLCP} we have plotted the evolution of electrostatic energy for the case of $L$ wave propagation of Case (D). In this case also similar behaviour is observed. 

\begin{figure*}
	\centering
	\includegraphics[width=6.0in]{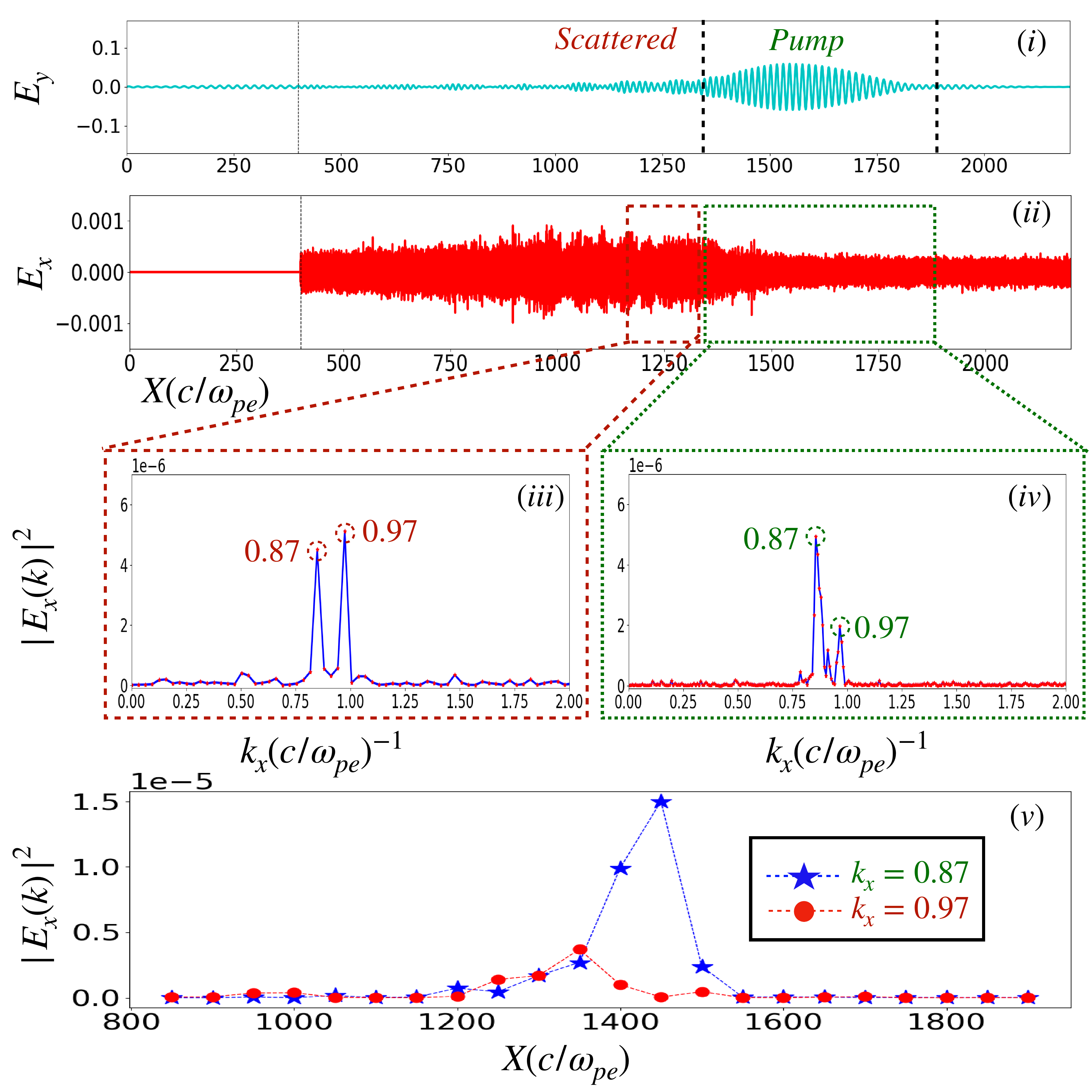}
	\caption{Figure shows the space Fast Fourier Transform (FFT) of the electrostatic field at various spatial block with respect to the pump wave at time $t = 3000$. Different subplots indicate the (i) Electromagnetic pulse both pump and scattered wave $E_y$, (ii) Electrostatic fluctuations $E_x$, FFT spectrum for different spatial blocks ((iii)  from $1200$ to $1400$, (iv) from $1400$ to $1800$), and (v) variation in power spectrum of electrostatic component $|E_x(k)|^2$ with the choice of spatial blocks.}
	\label{fig:PowerChangeExFFT}
\end{figure*}


We now discuss briefly the nonlinear phase of the instability. We choose a specific time of $t = 3000$ for the Case D at which the evolution of electrostatic energy is in its third phase,  showing a slow down in the growth. In Fig.\ref{fig:PowerChangeExFFT} we show the spatial fourier transform of the electrostatic field at various spatial blocks with respect to the pump pulse as illustrated in the figure. It can be observed that when the spatial FFT is taken just behind the pulse it shows a clear peak ($k_2 = 0.87$). At longer distances behind the pulse other peak in the spectra start appearing (e.g. $k= 0.97$). Thereafter it becomes broad  with multiple scales indicating the presence of nonlinear effects as expected. We have tracked the power of two different wave numbers by carrying out a Fourier transform in several blocks of localised spatial region and observe that while the power in $k = 0.87$ satisfying the Brillouin condition maximizes in the spatial regime 
overlapping with the pump pulse, the power in the second mode picks up thereafter. 

\section{Role of  smooth and sharp laser profiles}
\label{sec:LaserProfiles}
\begin{figure*}[ht]
	\centering
	\includegraphics[width=6.0in]{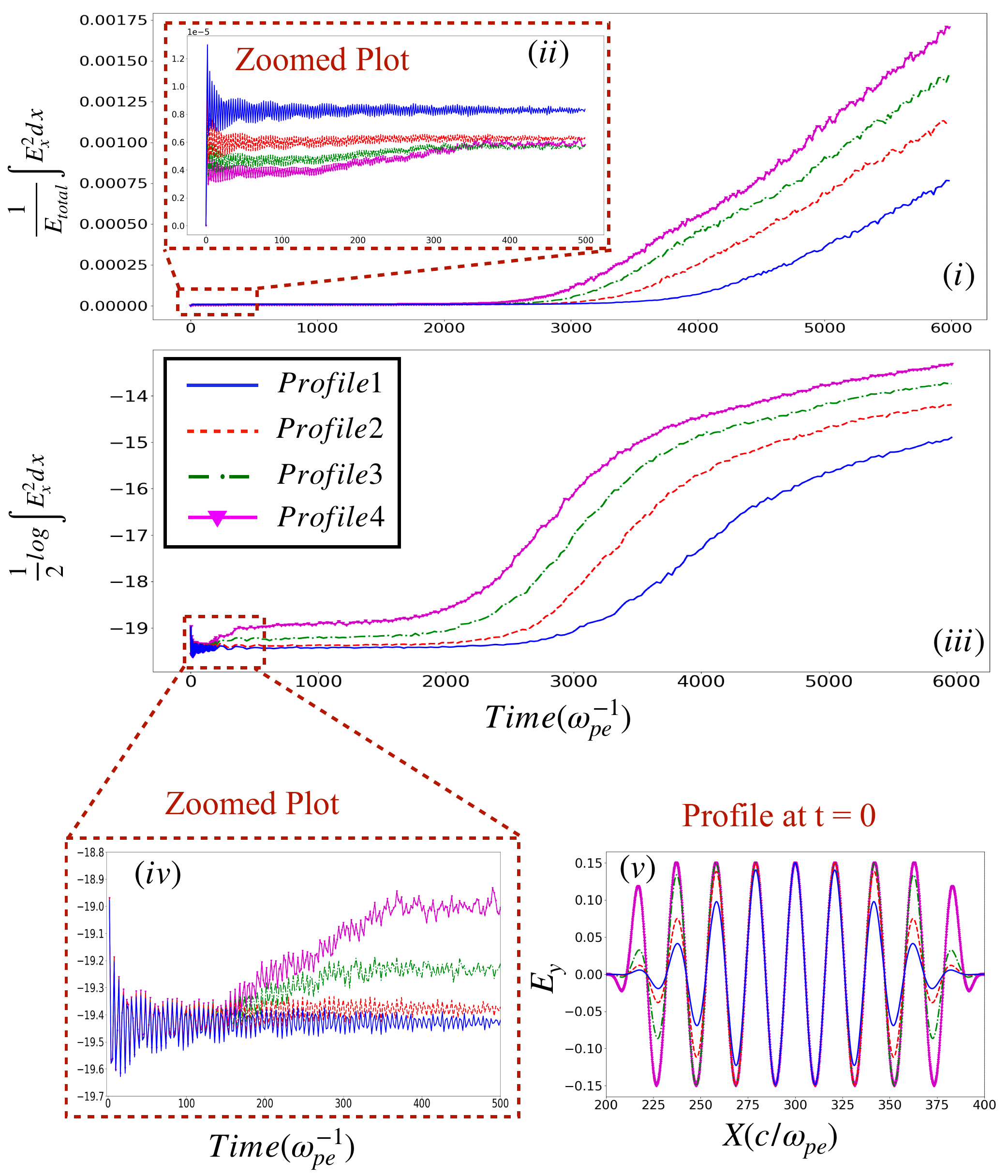}
	\caption{Figure highlights the effect of incident laser profile on the growth of Brillouin scattering process. Different subplots indicate (i) the evolution of electrostatic energy ((ii) zoomed plot) normalized by total input energy of incident laser, (iii) growth rate of Brillouin scattering for different laser profiles, and (v) four laser profiles used in simulation. It is evident from the plot that initially the ponderomotive pressure driven mechanism creates electrostatic fluctuations that are laser profile dependent and hence the Brillouin scattering mechanism occurs. }
	\label{fig:LaserProfileEffects}
\end{figure*}


In the work by Goswami et al. \cite{goswami2021ponderomotive}, published earlier for the same geometry a very sharp laser profile was chosen. There we observed the ponderomotive force driven electrostatic flluctutions. 
For the present studies we have chosen a smoother laser profile. In the section we describe a study where we carry out simulations with varying laser profile. The observations corresponding to four different laser profiles have been shown in Fig.\ref{fig:LaserProfileEffects} subplot(v). The sharpest profile has been denoted by the violet line with triangular dots. It can be observed from the subplot (iii) that there is an initial growth of electrostatic energy which is highest for this particular case. A zoomed picture has been shown in subplot(iv) of this region which also confirms the same. This is the time regiem where the initial ponderomotive pressure driven electrostatic fluctations occur. The thermalization of the effective electron kinetic energy acquired by this process leads to the effective temperature for the Brillouin process to occur later.  The   onset of the Brillouin process occurs earlier for the cases of the fourth profile where this happens fastest.  
Since the total energy associated with the pulse for these profiles may slightly differ. We have also compared  the evolution of the electrostatic  energy normalised by the total input energy from the laser pulse. It can be observed that the normalised electrostatic energy in the case of profile 4 is  higher compared to all other profiles after the scattering process takes place.

\section{Conclusion}
\label{sec:Conclusion}
With the help of OSIRIS-4.0, we have carried out a  comprehensive PIC simulations studies 
to understand the interaction of a laser pulse propagating in an overdense magnetize plasma 
for which the external magnetic field is directed  along the laser propagation direction. It has been shown that the laser energy can get  converted to the electrostatic energy by two distinct processes. The 
ponderomotive force driven process occurs first and is obviously found to be dependent on the sharpness of laser profile.  Such an excitation process was also shown  in an earlier publication by us \cite{goswami2021ponderomotive} where purposely  a very sharp laser pulse profile was chosen. 
Here we provide evidence of the Brillouin scattering process taking place. We also observe that there is a 
synergy between the ponderomotive  process leading to charge density fluctuations and the Brillouin back scattering process. The Brillouin process occurs subsequently and the timing of its  onset is related to 
electrostatic energy gained during the first process. 

\section{Acknowledgment} 
The authors would like to acknowledge the OSIRIS Consortium, consisting of UCLA ans IST
(Lisbon, Portugal) for providing access to the OSIRIS4.0 framework which is the work supported by NSF
ACI-1339893. This research work has been supported by
the Core Research Grant No. CRG/2018/000624 of Department
of Scient and Technology (DST), Government of India.
We also acknowledge support from J C Bose Fellowship Grant
of A D (JCB-000055/2017) from the Science and Engineering
Research Board (SFERB), Government of India. The authors
thank IIT Delhi HPC facility for computational resources.
Laxman thanks the Council for Scientific and Industrial
Research (Grant no. -09/086(1442)/2020-EMR-I) for funding
the research.

\section{References}
\bibliographystyle{unsrt}

\begin{thebibliography}{10}

\bibitem{das2020laser}
Amita Das.
\newblock Laser plasma session: Aapps-dpp conference, 12--17 nov 2018,
  kanazawa.
\newblock {\em Reviews of Modern Plasma Physics}, 4(1):1--16, 2020.

\bibitem{tochitsky2016prospects}
Sergei Tochitsky, Frederico Fiuza, and Chan Joshi.
\newblock Prospects and directions of co2 laser-driven accelerators.
\newblock In {\em AIP Conference Proceedings}, volume 1777, page 020005. AIP
  Publishing LLC, 2016.

\bibitem{haberberger2010fifteen}
D~Haberberger, S~Tochitsky, and C~Joshi.
\newblock Fifteen terawatt picosecond co 2 laser system.
\newblock {\em Optics express}, 18(17):17865--17875, 2010.

\bibitem{beg1997study}
FN~Beg, AR~Bell, AE~Dangor, CN~Danson, AP~Fews, ME~Glinsky, BA~Hammel, P~Lee,
  PA~Norreys, and Ma~Tatarakis.
\newblock A study of picosecond laser--solid interactions up to 1019 w cm- 2.
\newblock {\em Physics of plasmas}, 4(2):447--457, 1997.

\bibitem{nakamura2018record}
Daisuke Nakamura, A~Ikeda, H~Sawabe, YH~Matsuda, and S~Takeyama.
\newblock Record indoor magnetic field of 1200 t generated by electromagnetic
  flux-compression.
\newblock {\em Review of Scientific Instruments}, 89(9):095106, 2018.

\bibitem{korneev2015gigagauss}
Ph~Korneev, E~d'Humi{\`e}res, and V~Tikhonchuk.
\newblock Gigagauss-scale quasistatic magnetic field generation in a
  snail-shaped target.
\newblock {\em Physical Review E}, 91(4):043107, 2015.

\bibitem{goswami2021ponderomotive}
Laxman~Prasad Goswami, Srimanta Maity, Devshree Mandal, Ayushi Vashistha, and
  Amita Das.
\newblock Ponderomotive force driven mechanism for electrostatic wave
  excitation and energy absorption of electromagnetic waves in overdense
  magnetized plasma.
\newblock {\em Plasma Physics and Controlled Fusion}, 63(11):115003, 2021.

\bibitem{vashistha2020new}
Ayushi Vashistha, Devshree Mandal, Atul Kumar, Chandrasekhar Shukla, and Amita
  Das.
\newblock A new mechanism of direct coupling of laser energy to ions.
\newblock {\em New Journal of Physics}, 22(6):063023, 2020.

\bibitem{vashistha2021excitation}
Ayushi Vashistha, Devshree Mandal, and Amita Das.
\newblock Excitation of lower hybrid and magneto-sonic perturbations in laser
  plasma interaction.
\newblock {\em Nuclear Fusion}, 61(2):026016, 2021.

\bibitem{maity2021harmonic}
Srimanta Maity, Devshree Mandal, Ayushi Vashistha, Laxman~Prasad Goswami, and
  Amita Das.
\newblock Harmonic generation in the interaction of laser with a magnetized
  overdense plasma.
\newblock {\em Journal of Plasma Physics}, 87(5), 2021.

\bibitem{mandal2021transparency}
Devshree Mandal, Ayushi Vashistha, and Amita Das.
\newblock Electromagnetic wave transparency of x mode in strongly magnetized
  plasma.
\newblock {\em Scientific Reports}, 11(1), 2021.

\bibitem{kumar2019excitation}
Atul Kumar, Chandrasekhar Shukla, Deepa Verma, Amita Das, and Predhiman Kaw.
\newblock Excitation of kdv magnetosonic solitons in plasma in the presence of
  an external magnetic field.
\newblock {\em Plasma Physics and Controlled Fusion}, 61(6):065009, 2019.

\bibitem{mandal2020spontaneous}
Devshree Mandal, Ayushi Vashistha, and Amita Das.
\newblock Spontaneous formation of coherent structures by an intense laser
  pulse interacting with overdense plasma.
\newblock {\em Journal of Plasma Physics}, 86(6), 2020.

\bibitem{sano2020thermonuclear}
Takayoshi Sano, Shinsuke Fujioka, Yoshitaka Mori, Kunioki Mima, and Yasuhiko
  Sentoku.
\newblock Thermonuclear fusion triggered by collapsing standing whistler waves
  in magnetized overdense plasmas.
\newblock {\em Physical Review E}, 101(1):013206, 2020.

\bibitem{sano2019ultrafast}
Takayoshi Sano, Masayasu Hata, Daiki Kawahito, Kunioki Mima, and Yasuhiko
  Sentoku.
\newblock Ultrafast wave-particle energy transfer in the collapse of standing
  whistler waves.
\newblock {\em Physical Review E}, 100(5):053205, 2019.

\bibitem{kaw2017nonlinear}
PK~Kaw.
\newblock Nonlinear laser--plasma interactions.
\newblock {\em Reviews of Modern Plasma Physics}, 1(1):1--42, 2017.

\bibitem{drake1974parametric}
James~F Drake, Predhiman~K Kaw, Yee-Chun Lee, G~Schmid, Chuan~S Liu, and
  Marshall~N Rosenbluth.
\newblock Parametric instabilities of electromagnetic waves in plasmas.
\newblock {\em The Physics of Fluids}, 17(4):778--785, 1974.

\bibitem{liu1974raman}
CS~Liu, Marshall~N Rosenbluth, and Roscoe~B White.
\newblock Raman and brillouin scattering of electromagnetic waves in
  inhomogeneous plasmas.
\newblock {\em The Physics of Fluids}, 17(6):1211--1219, 1974.

\bibitem{saxena2007stability}
Vikrant Saxena, Amita Das, Sudip Sengupta, Predhiman Kaw, and Abhijit Sen.
\newblock Stability of nonlinear one-dimensional laser pulse solitons in a
  plasma.
\newblock {\em Physics of plasmas}, 14(7):072307, 2007.

\bibitem{sundar2011relativistic}
Sita Sundar, Amita Das, Vikrant Saxena, Predhiman Kaw, and Abhijit Sen.
\newblock Relativistic electromagnetic flat top solitons and their stability.
\newblock {\em Physics of Plasmas}, 18(11):112112, 2011.

\bibitem{kruer1973instability}
WL~Kruer, KG~Estabrook, and KH~Sinz.
\newblock Instability-generated laser reflection in plasmas.
\newblock Technical report, Univ. of California, Livermore, 1973.

\bibitem{tsytovich1973one}
VN~Tsytovich, L~Stenflo, H~Wilhelmsson, HG~Gustavsson, and K~Ostberg.
\newblock One-dimensional model for nonlinear reflection of laser radiation by
  an inhomogeneous plasma layer.
\newblock {\em Physica Scripta}, 7(5):241, 1973.

\bibitem{weiland1977coherent}
Jan Weiland and Hans Wilhelmsson.
\newblock Coherent non-linear interaction of waves in plasmas.
\newblock {\em Oxford Pergamon Press International Series on Natural
  Philosophy}, 88, 1977.

\bibitem{panwar2009stimulated}
Anuraj Panwar and AK~Sharma.
\newblock Stimulated brillouin scattering of the beat wave of two lasers in a
  plasma.
\newblock {\em Journal of Applied Physics}, 106(6):063301, 2009.

\bibitem{stenflo1990stimulated}
L~Stenflo.
\newblock Stimulated scattering of large amplitude waves in the ionosphere.
\newblock {\em Physica Scripta}, 1990(T30):166, 1990.

\bibitem{stenflo1995theory}
L~Stenflo.
\newblock Theory of stimulated scattering of large-amplitude waves.
\newblock {\em Journal of plasma physics}, 53(2):213--222, 1995.

\bibitem{shukla2010stimulated}
PK~Shukla and Lennart Stenflo.
\newblock Stimulated brillouin scattering of electromagnetic waves in
  magnetized plasmas.
\newblock {\em Journal of plasma physics}, 76(6):853--855, 2010.

\bibitem{jaiman1998stimulated}
NK~Jaiman and VK~Tripathi.
\newblock Stimulated brillouin scattering of an electromagnetic wave in a
  strongly magnetized plasma.
\newblock {\em Physics of Plasmas}, 5(1):222--226, 1998.

\bibitem{dawson1983particle}
John~M. Dawson.
\newblock Particle simulation of plasmas.
\newblock {\em Reviews of Modern Physics}, 55(2):403–447, 1983.

\bibitem{birdsall1991particle}
Charles~K Birdsall.
\newblock Particle-in-cell charged-particle simulations, plus monte carlo
  collisions with neutral atoms, pic-mcc.
\newblock {\em IEEE Transactions on plasma science}, 19(2):65--85, 1991.

\bibitem{hemker2000particle}
Roy~Gerrit Hemker.
\newblock {\em Particle-in-cell modeling of plasma-based accelerators in two
  and three dimensions}.
\newblock University of California, Los Angeles, 2000.

\bibitem{fonseca2002osiris}
Ricardo~A Fonseca, Luis~O Silva, Frank~S Tsung, Viktor~K Decyk, Wei Lu, Chuang
  Ren, Warren~B Mori, S~Deng, S~Lee, T~Katsouleas, et~al.
\newblock Osiris: A three-dimensional, fully relativistic particle in cell code
  for modeling plasma based accelerators.
\newblock In {\em International Conference on Computational Science}, pages
  342--351. Springer, 2002.

\bibitem{fonseca2008one}
RA~Fonseca, SF~Martins, LO~Silva, JW~Tonge, FS~Tsung, and WB~Mori.
\newblock One-to-one direct modeling of experiments and astrophysical
  scenarios: pushing the envelope on kinetic plasma simulations.
\newblock {\em Plasma Physics and Controlled Fusion}, 50(12):124034, 2008.

\end{thebibliography}

\end{document}